\documentstyle[12pt]{article}

\newcommand{\sect}[1]{\setcounter{equation}{0}\section{#1}}

\renewcommand{\thesection}{\arabic{section}}
\renewcommand{\thefootnote}{\fnsymbol{footnote}}
\newcommand{\bea}{\begin{eqnarray}}
\newcommand{\ena}{\end{eqnarray}}
\newcommand{\vs}[1]{\vspace{#1 mm}}

\renewcommand{\a}{\alpha}
\renewcommand{\b}{\beta}
\renewcommand{\c}{\gamma}

\newcommand{\e}{\epsilon}
\newcommand{\p}{\pi}
\newcommand{\n}{\nu}
\renewcommand{\t}{\tau}
\newcommand{\z}{\omega}
\newcommand{\G}{\Gamma}

\newcommand{\PR}[1]{Phys.\ Rev.\ {\bf #1}}

\newcommand{\PTP}[1]{Prog.\ Theor.\ Phys.\ {\bf #1}}
\newcommand{\AJ}[1]{Astorophys. \ J.\ {\bf #1}}
\newcommand{\JMP}[1]{J.\ Math.\ Phys.\ {\bf #1}}

\begin{document}
\topmargin 0pt
\oddsidemargin 5mm

\begin{titlepage}
\setcounter{page}{0}
\begin{flushright}
OU-HET 238\\
gr-qc/9603020\\
March,1996
\end{flushright}
\vs{5}
\begin{center}
{\Large{\bf Analytic Solutions of the Teukolsky Equation \\
            and their Low Frequency Expansions}}\\
\vs{8}
{\large  
Shuhei Mano,\footnote{e-mail address: mano@phys.wani.osaka-u.ac.jp}
Hisao Suzuki\footnote{e-mail address: hsuzuki@particle.phys.hokudai.ac.jp}
and Eiichi Takasugi\footnote{e-mail address: 
takasugi@phys.wani.osaka-u.ac.jp}
}\\
\vs{8}
{\em Department of Physics,
Osaka University \\ Toyonaka, Osaka 560, Japan} \\
{\em Department of Physics,
Hokkaido  University \\  Sapporo 060, Japan\dag} \\
\end{center}
\vs{8}
\centerline{{\bf Abstract}}
  Analytic solutions of the Teukolsky equation in Kerr geometries 
are presented in the form of series of hypergeometric functions 
and Coulomb wave functions. Relations between these solutions 
are established. 
The solutions provide a very  powerful method not only for examining  
the general properties of solutions and physical quantities 
when they are applied to, but also  for numerical computations. 
The solutions are given in the expansion of a small parameter 
$\e \equiv 2M\z$,   
$M$   being the mass of black hole, which  corresponds to 
Post-Minkowski expansion  by $G$ and to 
post-Newtonian expansion when they are applied to 
the gravitational radiation 
from a particle in circular orbit around a black hole. 
It is expected that these solutions will become a powerful 
weapon to construct the theoretical template towards 
LIGO and VIRGO projects.

\end{titlepage}

\newpage
\renewcommand{\thefootnote}{\arabic{footnote}}
\setcounter{footnote}{0}

\sect{Introduction}
\indent

There are growing interests in analytic solutions of the Teukolsky 
equation\cite{Teukolsky} in the Schwarzshild and the Kerr geometries 
in the connection 
of the gravitational wave cosmology. 
Since Teukolsky proposed the master equation for massless fields in the 
Kerr metrics, many efforts have been made to obtain the 
analytic solutions. The analytic expressions valid 
for low frequencies were 
found by Page\cite{Page}, Starobinsky and Churilov\cite{SC}  by matching the 
approximate 
solutions valid near horizon and far from it.  
Leaver\cite{Leaver}  made the systematic study to obtain the analytic 
solutions of Teukolsky equation in the form of series of 
various functions.  He found  the solution  
 in the form of series of Coulomb wave functions which is 
 valid in the region far from the horizon and established the relation between 
that solution and the one  in the form of the  Jaffe  type 
series which is valid near the horizon.  

Recently, Tagoshi and Nakamura\cite{TN} determined  
numerically the coefficients of the post-Newtonian expansion 
 of the gravitational radiation 
 by a particle traveling a circular orbit  around a 
Schwarzshild black hole.  Sasaki\cite{Sasaki}   
proposed  a method of post-Newtonian expansion to solve the 
homogeneous Regge-Wheeler equation  
by using   Bessel functions. 
Subsequently, the extensive study on this line was made by Tagoshi 
and Sasaki\cite{TS} and the result was compared with the one by Tagoshi 
and Nakamura.  The application of this method to the Kerr geometries 
was made by Shibata, Sasaki, Tagoshi and Tanaka\cite{SSTT}. 
Various other  applications  
was discussed by Poisson and Sasaki\cite{PS}. Now  the problem to obtain the 
analytic solutions and the 
examination of their  behaviors in low frequencies  became the  
important and the urgent topic. 

In this paper, we report that we  obtained  
 the analytic solutions of Teukolsky equation in Kerr geometries 
in the form of series of hypergeometric functions and Coulomb wave 
functions. The series solution of hypergeometric type is shown to 
be convergent 
in the region except infinity, while the one of Coulomb type is 
convergent in the region $\mid x \mid >1$, where 
$x=(r_+-r)/2M\sqrt{1-(a/M)^2}$ with $r_+$, $M$ and $a$ being 
the position of the outer horizon, the mass and the angular 
momentum of Kerr black hole. We established the relation between 
 two solutions with 
different regions of convergences.  
The solutions are interesting not only for 
the investigation of  general properties of solutions 
as mathematical physics, but also are for various applications 
to the gravitational wave cosmology. 
 The solutions are essentially given in the 
$\e$ expansion which corresponds to the Post-Minkowskian 
$G$ expansion   and also corresponds to 
the post-Newtonian expansion when they are applied to 
the problem of the gravitational radiation from a 
particle in cicular orbit  around a black hole so that 
our solutions are quite powerful to examine the $\e$ 
behaviors of various physical quantities. 
Our solutions are expected to become 
a powerful machine for numerical computation also 
because the convergences of series are well known.
Thus the solutions will become 
a powerfull weapons for the  construction of  the theoretical template 
towards the gravitatinal wave observation by LIGO and VIRGO 
projects. 

Our work was 
motivated by Sasaki's work.  We tried to improve  his  
method for  the solution of Regge-Wheeler equation  
because his method has several disadvantages: (1) it is difficult 
 to obtain the higher order terms of $\e \equiv 2M\z$, 
where $M$ and $\z$ being the Black hole mass and the angular 
frequency, (2) the expansion is not really the Bessel expansion 
because  coefficients are also variable dependent and 
(3) the convergence of the series was unknown. In order to 
improve these difficulties, we considered the solution 
in the form  of series of hypergeometric functions
 for the solutions of Regge-Wheeler 
equation and also for the Teukolsky equation in Schwarzshild metrics 
 and showed that  
the coefficients of series can be determined systematically 
in the expansion of $\e$ due to the recurrence relations 
among hypergeometric functions which we found\cite{MST}. 
This solution is valid near the horizon and not at infinity 
so that away from the horizon we have to consider the solution 
in the form of series of Coulomb wave function which was 
found by Leaver\cite{Leaver}. 
By matching these two solutions in the intermediate region, we obtained 
a good solution in the entire regions. 
After finishing our work, we happened to see the paper by Otchik\cite{Otchik}  
 who discussed 
the analytic solutions of the Teukoksky equation in the form of 
series of hypergeometric functions and Coulomb wave functions. 
We found that our method is essentially identical to Otchik's method, 
 but our solutions  disagreed with his ones. We compared 
our solutions and his ones and found that although various formulas which 
he presented were incorrect,  his  story itself turns out to be true. 
Since our results are  all different from those by Otchik\cite{Otchik} and 
the results themselves are quite important for the application, we 
present all results in this paper. 

We start from the Teukolsky equation which is separated  by writing 
\bea
\psi =e^{-i\omega t} e^{im \phi}S_{l}^m (\theta) R_{\z l m}(r).
\ena
The equation for $R$ is
\bea
&&\Delta R''+ 
{2(r-M)(s+1)}R'+\left[{{K^2-2is(r-M)K}\over{{\Delta}}}+4is\z r 
-\lambda \right]R=0,\nonumber\\
\ena
where $M$ is the mass of the black hole, $aM$ its angular momentum, 
$\Delta= r^2-2Mr+a^2 =(r-r_+)(r-r_-)$ with $r_\pm = M \pm \sqrt{M^2-a^2}$ 
where $r_+$ and $r_-$ are positions of outer and inner horizons, 
respectively,  
 $K=(r^2+a^2)\z-am, \lambda=E-s(s+1)-2ma\z+a^2\z^2$. The function 
 $S_l^m$  is the spin weighted 
spheroidal harmonics which will be discussed in Appendix A 
with the eigenvalue 
$E$\cite{Fackerell} 
\bea
E&=&l(l+1)-{2s^{2}m\xi\over{l(l+1)}}+[H(l+1)-H(l)-1]\xi^{2}+O(\xi^{3}),
\ena
\bea
H(l)={2(l^2-m^2)(l^2-s^2)^2\over{(2l-1)l^3(2l+1)}},
\ena
where $\xi=a \omega$ and $l$ is the angular momentum which takes an 
integer or half-integer 
number which satisfies $l \ge \max(\mid m \mid, \mid s \mid)$.

In Sec.2, we give the discussion about how we arrive 
 the analytic solutions in terms of hypergeometric 
functions and discuss their properties. In Sec.3, the analytic solutions 
in terms of Coulomb wave functions are given following the work of 
Leaver. The relation between two solutions in two different convergence 
regions is established in Sec.4. The low frequency expansion of 
these solutions is discussed in Sec.5. 
In Sec.6, the summary and remarks are given.

\indent

\sect{Analytic solution in the form of series of hypergeometric functions}
\indent
 
The radial Teukolsky equation has two regular singularities at $r=r_{\pm}$ and 
an irregular singularity at $r=\infty$. In order to obtain the solution 
in the form of series of hypergeometric functions, we have to deal with 
these  
regular singularities. Following the 
 discussion in Appendix B, we take the form of $R$ which satisfies the 
incoming 
boundary condition on the outer horizon. In particular, we choose the form 
given by $(\alpha_-,\beta_+)$ in the 
notation in Appendix B with the variable $x=\z (r_+ -r)/\e \kappa$ as
\bea
R_{in}^{\n}&=&e^{i\e\kappa x}(-x)^{-s-i{\e+\t\over{2}}}
(1-x)^{i{\e-\t\over{2}}}p_{in}^{\n}(x),
\ena
where $\e=2M\z, q={a\over{M}}, \kappa={\sqrt{1-q^2}}$ and 
$\t={{\e-mq}\over{\kappa}}$.\\
Then, the radial Teukolsky equation becomes
\bea
&&x(1-x)p_{in}^{\n}{}''+[1-s-i\e-i\t-(2-2i\t)x]p_{in}^{\n}{}'
+(\n+i\t)(\n+1-i\t)p_{in}^{\n}\nonumber\\
&=&2i\e \kappa[-x(1-x)p_{in}^{\n}{}'+(1-s+i\e-i\t)xp_{in}^{\n}]\nonumber\\
&+&[-\lambda-s(s+1)+\n(\n+1)+\e^2-i\e \kappa(1-2s)]p_{in}^{\n},
\ena
Here we introduced  the parameter $\n$ as the renormalized angular momentum 
which satisfies $\n=l +O(\e)$. Then, the right-hand side of Eq.(2.2) is 
of order $\e$ so that this form of equation is suitable to obtain 
the solution in the expansion of $\e$. The zeroth order solution of 
Eq.(2.2) is the hypergeometric function. 

From the structure of the above equation, the solution 
may  be written in the form of series of hypergeometric functions as
\bea
p_{in}^{\n}(x)=\sum_{n=-\infty}^{\infty}a_{n}^{\n}p_{n+\n}(x),
\ena
where 
\bea
p_{n+\n}(x)&=&F(n+\nu+1-i\t,-n-\n-i\t;1-s-i\e-i\t;x),
\ena
with the use of the renormalized angular momentum $\n$ rather than 
$l$. We expect that the series will coincide to the $\e$ expansion. 
In order for the coeficients of  series (2.3) to be a solved, 
it is essential that 
 the coefficients $a_n^{\n}$'s satisfy the three term recurrence relation. 
For this, terms such as $x(1-x){p'}_{n+\n}$ and $xp_{n+\n}$ must be 
expressed as linear combinations of $p_{n+\n+1}$, $p_{n+\n}$ 
and $p_{n+\n-1}$. Amazingly enough, we found the following recurrence 
relations,
\bea
xp_{n+\n}&=&-{(n+\n+1-s-i\e)(n+\n+1-i\t)\over{2(n+\n+1)(2n+2\n+1)}}p_{n+\n+1}
\nonumber\\
&+&{1\over{2}}\left[1+{i\t(s+i\e)\over{(n+\n)(n+\n+1)}}\right]
p_{n+\n}\nonumber\\
&-&{(n+\n+s+i\e)(n+\n+i\t)\over{2(n+\n)(2n+2\n+1)}}p_{n+\n-1},
\ena

\bea
x(1-x)p'_{n+\n}&=&{(n+\n+i\t)(n+\n+1-i\t)(n+\n+1-s-i\e)
\over{2(n+\n+1)(2n+2\n+1)}}p_{n+\n+1}\nonumber\\
&+&{1\over{2}}(s+i\e)\left[1+{i\t(1-i\t)\over{(n+\n)(n+\n+1)}}\right]
p_{n+\n}\nonumber\\
&-&{(n+\n+1-i\t)(n+\n+i\t)(n+\n+s+i\e)\over{2(n+\n)(2n+2\n+1)}}p_{n+\n-1},
\ena 
which enables us to obtain the three term recurrence relation among 
$a_n^{\n}$. The above recurrence relations among hypergoemetric functions 
 can be proved by using the power series expansions. By substituting 
the form in Eq.(2.4) into the radial Teukolsky equation (2.2), 
we find that $p_{in}^{\n}$ becomes a solution 
if the following  recurrence relation is 
satisfied:
\bea
\a_n^{\n} a_{n+1}^{\n}+\b_n^{\n} a_n^{\n}+\c_n^{\n} a_{n-1}^{\n}=0,
\ena
where
\bea
\a_n^{\n}&=&{i\e \kappa (n+\n+1+s+i\e)(n+\n+1+s-i\e)(n+\n+1+i\t)
\over{(n+\n+1)(2n+2\n+3)}},
\ena
\bea
\b_n^{\n}&=&-\lambda-s(s+1)+(n+\n)(n+\n+1)+\e^2+\e(\e-mq)\nonumber\\
&+&{\e (\e-mq)(s^2+\e^2) \over{(n+\n)(n+\n+1)}},
\ena
\bea
\c_n^{\n}&=&-{i\e \kappa (n+\n-s+i\e)(n+\n-s-i\e)(n+\n-i\t)
\over{(n+\n)(2n+2\n-1)}}.
\ena

By introducing the continued fractions 
\bea
R_n (\n)={a_n^{\n}\over{a_{n-1}^{\n}}},
\qquad
L_n(\n)={a_n^{\n}\over{a_{n+1}^{\n}}},
\ena
we find 
\bea
R_n(\n)=-{\c_n^{\n}\over{\b_n^{\n}+\a_n^{\n}R_{n+1}(\n)}},
\qquad
L_n(\n)=-{\a_n^{\n}\over{\b_n^{\n}+\c_n^{\n}L_{n-1}(\n)}}.
\ena
From these equations, we can evaluate  the coefficients by taking 
the initial condition $a_0^{\n}=1$. The renormalized angular 
momentum $\n$ is determined by requiring that  
the coefficients obtained by using $R_n(\n)$ agree with 
those by using $L_n(\n)$, that is, by 
solving the transcnedental equation for $\n$ 
\bea
R_n(\n)L_{n-1}(\n)=1.
\ena
If Eq.(2.13) is satisfied, we find 
\bea
\displaystyle \lim_{n \rightarrow \infty} n\frac{a_n^{\n}}{a_{n-1}^{\n}}=
-\lim_{n \rightarrow -\infty}  n\frac{a_n^{\n}}{a_{n+1}^{\n}}
=\frac {i\e \kappa }2.
\ena
From the large $n$ behavior of hypergeometric functions, we find by using 
the recurrence formula of hypergeometric functions (2.5) as\cite{Bateman}
\bea
\displaystyle \lim_{n \rightarrow \infty} 
\frac{p_{n+\n}(x)}{p_{n+\n -1}(x)}=
\lim_{n \rightarrow -\infty}  
\frac{p_{n+\n}(x)}{p_{n+\n +1}(x)}
=1-2x +((1-2x)^2-1)^{1/2}.
\ena
From Eqs.(2.14) and (2.15), we find 
\bea
\displaystyle \lim_{n \rightarrow \infty} 
\frac{n a_n^{\n}p_{n+\n}(x)}{a_{n-1}^{\n}p_{n+\n -1}(x)}=
-\lim_{n \rightarrow -\infty}  
\frac{n a_n^{\n}p_{n+\n}(x)}{a_{n+1}^{\n}p_{n+\n +1}(x)}
=\frac {i\e \kappa }{2}[1-2x +((1-2x)^2-1)^{1/2}].
\ena
Thus the series converges in all the 
complex plane of $x$ except for $x=\infty$.

As for the recurrence relation (2.8), we find   
$\alpha_{-n}^{-\n-1}=\gamma_n^{\n}$ and   
$\gamma_{-n}^{-\n-1}=\alpha_n^{\n}$ so that  
 $a_{-n}^{-\n-1}$ satisfies the same recursion relation as  that  
which $a_{n}^{\n}$ does. Thus  if we choose 
$a_0^{\n}=a_0^{-\n-1}=1$,  we have 
\bea
a_n^{\n}=a_{-n}^{-\n-1}.
\ena
Also, we find 
\bea
R_n(-\n-1)L_{n-1}(-\n-1)=R_{-n+1}(\n)L_{-n}(\n)=1,
\ena
which means that if $\n$ is the solution of Eq.(2.13), then $-\n-1$ 
is also the solution. 

It is easily seen that the solution $R_{in}^{\n}$ is symmetric under the 
exchange of $\n$ with $-\n-1$ as follows.  By using the 
formula
\bea
p_{n+\n}(x)&=&\frac{\G (1-s-i\e-i\t)\G (2n+2\n+1)}
{\G (n+\n+1-i\t)\G (n+\n+1-s-i\e)}(-x)^{n+\n+i\t}\nonumber\\
&\times&F(-n-\n-i\t,-n-\n+s+i\e;-2n-2\n;\frac{1}{x})\nonumber\\
&+&\frac{\G (1-s-i\e-i\t)\G (-2n-2\n-1)}{\G (-n-\n-i\t)\G (-n-\n-s-i\e)}
(-x)^{-n-\n+i\t}\nonumber\\
&\times&F(n+\n+1-i\t,n+\n+1+s+i\e;2n+2\n+2;\frac{1}{x}),
\ena
we can show
\bea
R_{in}^{\n}=R_0^{\n}+R_0^{-\n-1},
\ena
where 
 \bea
R_0^{\n}&=&e^{i\e \kappa x}(-x)^{\n-s-{i\over{2}}(\e-\t)}
(1-x)^{{i\over{2}}(\e-\t)}\sum_{n=-\infty}
^{\infty}a_n^{\n}{\G(1-s-i\e-i\t)\G(2n+2\n+1)
\over{\G(n+\n+1-i\t)\G(n+\n+1-s-i\e)}}\nonumber\\
&\times&(-x)^{n}F(-n-\n-i\t,-n-\n+s+i\e;-2n-2\n;{1\over{x}}).
\ena

The behavior of $R_{in}^{\n}$  on the outer horizon
 $(x = 0)$ is
\bea
R_{in}^{\n}\rightarrow(-x)^{-s-{i\over{2}}(\e+\t)}
\sum_{n=-\infty}^{\infty}a_n^{\n},
\ena
which gives the normalization of our solution.

We can also show that $R_0^{\n}$ and $R_0^{-\n-1}$ are  solutions which 
are independent each other. To see this explicitly, 
we consider the solution which satisfies the outgoing boundary 
condition on the 
outer horizon. The outgoing solution which corresponding to 
$(\alpha_+,\beta_-)$ in the notation defined in Appendix B 
can be written as 
\bea
R_{out}^{\n}=e^{i\e \kappa x}(-x)^{{i\over{2}}(\e+\t)}
(1-x)^{-s-{i\over{2}}(\e-\t)}p_{out}^{\n}.
\ena
Now we expand $p_{out}^{\n}$ as 
\bea
p_{out}^{\n}(x)=\sum_{n=-\infty}^{\infty}\tilde{a_n^{\n}}\tilde{p}_{n+\n}(x),
\ena
where 
\bea
\tilde{p}_{n+\n}(x)=F(n+\n+1+i\t,-n-\n+i\t;1+s+i\e+i\t;x).
\ena
Similarly to the solution satisfying the incoming boundary condition,  
we find that the above series becomes a solution if the following 
recurrence relation is satisfied;
\bea
\tilde{\a}_n^{\n}\tilde{a}_{n+1}^{\n}+{\b}_n^{\n}\tilde{a}_n^{\n}
+\tilde{\c}_n^{\n}\tilde{a}_{n-1}^{\n}=0,
\ena
\bea
\tilde{\a}_n^{\n}={(n+\n+1-s-i\e)(n+\n+1-i\t)
\over{(n+\n+1+s+i\e)(n+\n+1+i\t)}}\a_n^{\n},
\ena
\bea
\tilde{\c}_n^{\n}={(n+\n+s+i\e)(n+\n+i\t)\over{(n+\n-s-i\e)(n+\n-i\t)}}
\c_n^{\n},
\ena
where $a_n^{\n}$, ${\b}_n^{\n}$ and $\c_n^{\n}$ are defined in 
Eqs.(2.8)-(2.10). 
By inspection, we see that this recurrence relation is reduced to  
the  one in Eq.(2.7) by redefing systematically the coeficients as 
\bea
\tilde{a_n^{\n}}={\G(\n+1-s-i\e)\G(\n+1-i\t)
\G(n+\n+1+s+i\e)\G(n+\n+1+i\t)\over{\G(\n+1+s+i\e)\G(\n+1+i\t)
\G(n+\n+1-s-i\e)\G(n+\n+1-i\t)}}a_n^{\n},
\ena
where we chose $\tilde a_0^{\n}=1$. 
Now we take $\tilde{a}_0^{-\n-1}=\tilde {a}_0^{\n}=1$, then 
after some computation we find 
\bea
R_{out}^{\n}=A_{\n}R_0^{\n}+A_{-\n-1}R_0^{-\n-1},
\ena
where
\bea
A_{\n}={\G(1+s+i\e+i\t)\G(\n+1-s-i\e)\G(\n+1-i\t)
\over{\G(1-s-i\e-i\t)\G(\n+1+s+i\e)\G(\n+1+i\t)}}.
\ena
This relation explicitly shows that $R_0^{\n}$ and $R_0^{-\n-1}$ are 
independent solutions of Eq.(2.1).

\sect{Analytic solutions in the form of series of Coulomb wave functions}  
\indent

 Analytic solution in the form of series of  Coulomb wave functions 
are given by Leaver\cite{Leaver}. Here, we follow the 
discussion in Appendix B and start the parameterization to 
remove the singularity at $r=r_-$.  By using 
a variable $z=\z (r-r_+) =-\e \kappa  x$, we take the following form
\bea
R_C^{\n}&=&z^{-1-s}\left(1+{\e \kappa \over{z}}\right)
^{{i\over{2}}(\e-\t)}f_{\n}(z).
\ena
Then, we find 
\bea
&&z^2f_{\n}''+[z^2+2(\e +is)z-\n(\n+1)]f_{\n} \nonumber \\
&=&-\e \kappa z(f_{\n}''+f_{\n})+\e \kappa (1+s+i\e-i\t)f_{\n}'
-{\e \kappa (1+s+i\e)(1-i\t)\over{z}}f_{\n}\nonumber\\
&+&[\lambda+s(s+1)-\n(\n+1)-2\e^2 +\e mq -\e \kappa (\e+is)]f_{\n}.
\ena
If we consider $\n$ to be $\n=l+O(\e)$, the right-hand side of 
Eq.(3.2) is the quantity of order $\e$ so that this 
equation is a suitable one to obtain the solution in the 
expansion of $\e$. 

Here we aim to obtain the exact solution by expanding  $f_{\n}(z)$ 
in terms of Coulomb functions with the renormalized angular momentum 
$\n$,
\bea
f_{\n}=\sum_{n=-\infty}^{\infty}b_n^{\n}F_{n+\n}(z),
\ena
where $F_{n+\n}$ is the unnormalized Coulomb wave function, 
\bea
F_{n+\n}=e^{-iz}(2z)^{n+\n}z
{\Gamma(n+\n+1-s+i\e)\over{\Gamma(2n+2\n+2)}}
\Phi(n+\n+1-s+i\e,2n+2\n+2;2iz).
\ena  
It is essential for the solution  of Coulomb wave functions to be 
related to the one of hypergeometric functions, 
the renormalized angular momentum $\n$ takes the same 
value for both cases.

By substituting Eq.(3.3) into Eq.(3.2) and using the 
recurrence relations satisfied by the Coulomb wave functions,
\bea
{1\over{z}}F_{n+\n}&=&{(n+\n+1+s-i\e)\over{(n+\n+1)(2n+2\n+1)}}F_{n+\n+1}
+{is+\e\over{(n+\n)(n+\n+1)}}F_{n+\n}\nonumber\\
&+&{(n+\n-s+i\e)\over{(n+\n)(2n+2\n+1)}}F_{n+\n-1},
\ena
\bea
F'_{n+\n}&=&-{(n+\n)(n+\n+1+s-i\e)\over{(n+\n+1)(2n+2\n+1)}}F_{n+\n+1}
+{is+\e\over{(n+\n)(n+\n+1)}}F_{n+\n}\nonumber\\
&+&{(n+\n+1)(n+\n-s+i\e)\over{(n+\n)(2n+2\n+1)}}F_{n+\n-1},
\ena  
we obtain the three term recursion relation of coefficients $b_n^{\n}$,
\bea
{\a'}_n^{\n}b_{n+1}^{\n}+\b_n^{\n}b_n^{\n}+{\c'}_n^{\n}b_{n-1}^{\n}=0,
\ena
\bea
{\a'}_n^{\n}=-i{(n+\n+1-s+i\e)(n+\n+1-s-i\e)
\over{(n+\n+1+s+i\e)(n+\n+1+s-i\e)}}\a_n^{\n}
\ena
\bea
{\c'}_n^{\n}=i{(\n+n+s+i\e)(n+\n+s-i\e)\over{(n+\n-s+i\e)(n+\n-s-i\e)}}\c
_n^{\n},
\ena
where $\a_n^{\n}$, ${\b}_n^{\n}$ and $\c_n^{\n}$ are defined in 
Eqs.(2.8)-(2.10). By inspection, we see that 
this recurrence relation is deformed to   
the  one in Eq.(2.7) if we systematically redefine the  coefficients  as 
\bea
b_n^{\n}=i^n{\G(\n+1-s+i\e)\G(\n+1-s-i\e)\G(n+\n+1+s+i\e)\G(n+\n+1+s-i\e)
\over{\G(\n+1+s+i\e)\G(\n+1+s-i\e)\G(n+\n+1-s+i\e)\G(n+\n+1-s-i\e)}}a_n^{\n},
\nonumber\\
\ena
where we chose the initial condition  $b_0^{\n}=1$. 
Since the recurrence relation obtained for the Coulomb expansion 
case is identical to  the one for the hypergeometric case,  
the renormalized angular momenta $\n$ derived from  both 
solutions are the same which  allows us to relate these two solutions. 

As for the convergence of series in Eq.(3.3), we find 
\bea
\displaystyle \lim_{n \rightarrow \infty} n\frac{b_n^{\n}}{b_{n-1}^{\n}}=
\lim_{n \rightarrow -\infty}  n\frac{b_n^{\n}}{b_{n+1}^{\n}}
=-\frac {\e \kappa }2,
\ena
and from the recurrence relation (3.5) 
\bea
\displaystyle \lim_{n \rightarrow \infty} 
\frac{F_{n+\n}(z)}{nF_{n+\n-1 }(z)}=
\lim_{n \rightarrow -\infty}  
\frac{F_{n+\n}(z)}{nF_{n+\n +1}(z)}
=\frac{2}{z},
\ena
so that 
\bea
\displaystyle \lim_{n \rightarrow \infty} 
\frac{b_n^{\n}F_{n+\n}(z)}{b_{n-1}^{\n}F_{n+\n-1 }(z)}=
\lim_{n \rightarrow -\infty}  
\frac{b_n^{\n}F_{n+\n}(z)}{b_{n+1}^{\n}F_{n+\n +1}(z)}
=-\frac{\e \kappa }{z},
\ena
Thus we find  that the series converges for $z > \e \kappa  $ or 
$\mid x\mid  > 1$.

In order to  derive the asymptotic behavior 
of the Coulomb solution $R_C^{\n}$, it is useful to rewrite  
as
\bea
R_C^{\n}=R_{C\,in}^{\n}+R_{C\,out}^{\n}
\ena
where 
\bea
R_{C\,in}^{\n}&&=e^{-iz}z^{\n-s}\left(1+{\e \kappa \over{z}}\right)
^{{i\over{2}}(\e-\t)}2^{\n}e^{i\p(\n+1-s+i\e)}\nonumber\\
&\times&\sum_{n=-\infty}^{\infty}b_n^{\n}(-2z)^n{\G(n+\n+1-s+i\e)
\over{\G(n+\n+1+s-i\e)}}\Psi(n+\n+1-s+i\e,2n+2\n+2;2iz).\nonumber\\
\ena

\bea
R_{C\,out}^{\n}&=&e^{iz}z^{\n-s}\left(1+{\e \kappa \over{z}}\right)^
{{i\over{2}}(\e-\t)}2^{\n}e^{-i\p(\n+1+s-i\e)}\nonumber\\
&\times&\sum_{n=-\infty}^{\infty}b_n^{\n}(-2z)^n
\Psi(n+\n+1+s-i\e,2n+2\n+2;-2iz),
\ena

Another independent solution is obtained by replacing $\n$ with $-\n-1$ 
because $-\n-1$ is the renormalized angular momentum if $\n$ is the one. 
 Thus, we have  another independent 
solution   by $R_C^{-\n-1}$. The coefficients  $b_{-n}^{-\n-1}$ are 
obtained from  $b_n^{\n}$ by the relation  
\bea
b_{-n}^{-\n-1}=(-1)^n b_n^{\n} 
\ena
by choosing  $b_0^{\n}=b_0^{-\n-1}=1$ in conformity with Eq.(3.10). 
With the use of Eq.(3.14), we find 
by using the identity $\Psi(-L\pm s \mp i\e,-2L,x)=
x^{2L+1}\Psi (L+1\pm s \mp i\e,2L+2,x)$,
\bea
R_{C in}^{-\n-1}&=&-ie^{-i\pi \n}\frac{\sin \pi(\n-s+i\e)}
{\sin \pi(\n+s-i\e)}R_{C in }^{\n}, \\
R_{C out}^{-\n-1}&=&ie^{i\pi \n}R_{C out }^{\n}.
\ena
Thus the solution $R_C^{-\n-1}$ is expressed by 
\bea
R_C^{-\n-1}= -ie^{-i\pi \n}\frac{\sin \pi(\n-s+i\e)}
{\sin \pi(\n+s-i\e)}R_{C in }^{\n} +ie^{i\pi \n}R_{C out }^{\n}.
\ena

\sect{The relation between two solutions}

First we notice that $R_0^{\n}$ and $R_C^{\n}$ are solutions of Teukolsky 
equation. Second we see that if we expand  
 these solutions in  Laurent series of $x=-z/\e \kappa $,  
both solutions give the series with   
the same characteristic exponent at 
$x \rightarrow \infty $. Thus, $R_0^{\n}$ must be proportional 
to  $R_C^{\n}$,
\bea
R_0^{\n}=K_{\n}R_C^{\n}.
\ena
The constant factor  $K_{\n}$ is determined  
 by comparing  like terms of these  series. 
We find 
\bea
K_{\n}&=&{(\e \kappa )^{-\n-r+s}2^{-\n-r}(-i)^r\G(1-s-i\e-i\t)
\over{\G(1+r+\n+i\t)\G(1+r+\n-s-i\e)\G(1+r+\n-s+i\e)}}\nonumber\\
&\times&\sum_{n=r}^{\infty}
{\G(n+\n+1+i\t)\G(n+r+2\n+1)\over{(n-r)!\G(n+\n+1-i\t)}}a_n^{\n}\nonumber\\
&\times&[{\sum_{n=-\infty}^{r}
(-i)^n{b_n^{\n}\over{(r-n)!\G(n+r+2\n+2)}}}]^{-1},\nonumber\\
\ena
where $r$ is an arbitrary integer.

By using these relations,  $R_{in}^{\n}$ can be written by 
using the Coulomb expansion solutions as
\bea
R_{in}^{\n}&=&(K_{\n}R_{C\,in}^{\n}+K_{-\n-1}R_{C\,in}^{-\n-1})
+(K_{\n}R_{C\,out}^{\n}+K_{-\n-1}R_{C\,out}^{-\n-1})
\nonumber\\
&=&
(K_{\n}-ie^{-i\pi \n}\frac{\sin \pi(\n-s+i\e)}
{\sin \pi(\n+s-i\e)}K_{-\n-1})R_{C in }^{\n}
 +(K_{\n}+ie^{i\pi \n}K_{-\n-1})R_{C out }^{\n}.
\ena
The asymptotic behavior at $z \rightarrow \infty $ is 
\bea
R_{in}^{\n}&=&A_{out}^{s\,\n}e^{iz}z^{-2s-1+i\e}+A_{in}^{s\,\n}e^{-iz}
z^{-1-i\e},
\ena
where $A_{out}^{s\,\n}$ and $A_{in}^{s\,\n}$ are amplitudes of the 
outgoing and incoming waves at infinity  of  the solution which satisfys the 
incoming boundary condition at the outer horizon. They are given by 
\bea
A_{out}^{s\,\n}&=&e^{-{i\over{2}}\p (\n+1+ s-i\e)}2^{-1-s+i\e}(K_{\n}
+ie^{i\p\n}K_{-\n-1})\sum_{n=-\infty}^{\infty}b_n^{\n}(-i)^n,
\ena
and
\bea
A_{in}^{s\,\n}&=&e^{-{i\over{2}}\p(-\n-1+s-i\e)}2^{-1+s-i\e}\left(K_{\n}-i
e^{-i\p\n}{\sin\p(\n-s+i\e)\over{\sin\p(\n+s-i\e)}}K_{-\n-1}\right)
\nonumber\\
&\times&\sum_{n=-\infty}^{\infty}b_n^{\n}i^n{\G(n+\n+1-s+i\e)
\over{\G(n+\n+1+s-i\e)}}.
\ena

One application of these amplitudes is to derive the  absorption coefficients. 
By using the method  given in Ref.\cite{TeukolskyPress}, the absorption 
coefficient $\G$ can be expressed in terms of $A_s^{in}$ and $A_s^{out}$ 
as follows;
\bea
\G^{s\,\n}=1-\left|{A_{out}^{-s\,\n}A_{out}^{s\,\n}
\over{A_{in}^{-s\,\n}A_{in}^{s\,\n}}}\right|
\ena

In the end of this section, we show how the upgoing solution 
which satisfies the outgoing boundary condition at infinity 
is expressed 
in terms of $R_0^{\n}$ and $R_0^{-\n-1}$ defined in Eq.(2.21). 
From Eq.(2.30), we find 
\bea
R_{out}^{\n}&=&
(A_{\n}K_{\n}-ie^{-i\pi \n}\frac{\sin \pi(\n-s+i\e)}
{\sin \pi(\n+s-i\e)}A_{-\n-1}K_{-\n-1})R_{C in }^{\n}\nonumber\\
 &&\hskip 2mm +(A_{\n}K_{\n}+ie^{i\pi \n}A_{-\n-1}K_{-\n-1})R_{C out }^{\n}.
\ena
By using Eqs.(4.3) and (4.8), we obtain
\bea
R_{up}^{\n}&=&R_{C out}^{\n}\nonumber\\
 &=& \left [ \frac{\sin \pi(\n-s+i\e)}
{\sin \pi(\n+s-i\e)}(K_{\n})^{-1}R_0^{\n}-ie^{i\pi \n}
(K_{-\n-1})^{-1}R_0^{-\n-1}   \right ]\nonumber\\
 && \hskip 2mm \times \left [ e^{2i\pi \n}+\frac{\sin \pi(\n-s+i\e)}
{\sin \pi(\n+s-i\e)}  \right ]^{-1}.
\ena

\sect{Low frequency  expansions of solutions}
 In this section, we discuss how to derive the solution in the expansion 
of the small parameter $\e=2M\z$. 
In order to find the solution in Eqs.(2.1) and (4.3) up to some 
power of $\e$, we have to calculate $\n$ and $a_n^{\n}$ to that 
order by using Eq.(2.12) and (2.13) with the condition (2.14) and 
$a_0^{\n}=a_0^{-\n-1}=1$. Other coefficients $b_n^{\n}$ can be calculated 
from $a_n^{\n}$ by using the formula (3.10).

For $a_n^{\n}$ with $n \ge 1$, the equation for $R_n(\n)$ is 
useful. Since   $\a_n^{\n}, \gamma_n^{\n}  \sim O(\e)$ and 
and $\b_n^{\n} \simeq n(n+2l+1) \sim O(1)$, we find 
$R_n(\n) \sim O(\e)$ for all positive integer $n$. 
As a result with $a_0^{\n}=1$, 
we find  
\bea
a_n^{\n} \sim O(\e^n)\hskip 1cm {\rm for} \hskip 5mm 
n\ge 1.
\ena

Before discussing the coeficients for $n<0$, 
 we  derive the renormalized angular 
momentum $\n$ up to $O(\e^2)$. For this, it is convenient to  use 
the constraint for $n=1$, $R_1(\n)L_0(\n)=1$. We notice that 
$R_1(\n)\sim O(\e)$ so that  $L_0(\n)$ must behave as $O(1/\e)$.  
which requires that $\b_0^{\n} + \gamma_0^{\n}L_{-1}(\n) \sim O(\e^2)$ 
because $\a_0^{\n} \sim O(\e)$. In order to obtain 
$\n$ up to $O(\e)$, we need to know 
 the information of $\b_0^{\n}$ up to $O(\e^2)$ where 
the second order term of $\n$ involves. Thus, we need the 
information about $R_1(\n)$, $L_{-1}(\n)$, $\a_{0}^{\n}$ and 
$\gamma_{0}^{\n}$ up to $O(\e)$. Here we assume that 
$L_{-2}(\n)\sim O(\e)$ whose validity will be discussed later. 
In this situation, $R_1(\n)$, $L_{-1}(\n)$, $\a_{0}^{\n}$ and 
$\gamma_{0}^{\n}$ can be calculated immediately.  
By substituting these to the constraint eqaution $R_1(\n)L_0(\n)=1$ 
to find 
\bea
\n&=& l+{1\over{2l+1}}\left[-2-{s^2\over{l(l+1)}}+{[(l+1)^2-s^2]^2
\over{(2l+1)(2l+2)(2l+3)}}-{(l^2-s^2)^2\over{(2l-1)2l(2l+1)}}\right]
\e^2+O(\e^3).\nonumber\\
\ena
The fact that the correction term of $\n$ starts from the second order term 
of $\e$ simplifies the calculation of the coeficients up to $O(\e^2)$. 

Now we discuss the coeficients for  negative integer $n$ 
for $s \neq 0$ which are 
derived by using  the equation for $L_n(\n)$. 
For large negative value of $\mid n \mid$, 
$L_n(\n) \simeq  -i\e \kappa /2n$. Most of the negative integer value of 
$n$, $L_n(\n) \sim O(\e)$. There arise some exceptions  
for certain values of $n$ because  
the denominator of $\a_n^{\n}$ vanishes  
at $n=-l-1$ or $ -l-\frac 32$ and  
also $\b_n^{\n}$ vanishes at $n=-2l-1$ in the zeroth order of $\e$. 
Because of this, we find for integer $l$'s, 
\bea
L_{-l-1}(\n) \sim O(1)&&, \nonumber\\
L_{-2l-1}(\n) \sim O(1/\e)&&, \nonumber\\
L_n(\n) \sim O(\e) && {\rm for \hskip 1mm all \hskip 1mm others}.
\ena
We also find for half-integer $n$'s,
\bea
L_{-(l+\frac 12)-1}(\n) \sim O(1/\e)&&, \nonumber\\
L_{-2(l+\frac 12)}(\n) \sim O(1/\e)&&, \nonumber\\
L_n(\n) \sim O(\e) && {\rm for \hskip 1mm all \hskip 1mm others}.
\ena

From the above estimates, we find 
for a integer $l$, 
\bea
a_n^{\n} &\sim& O({\e}^{\mid n\mid}), \hskip 5mm {\rm for } \hskip 5mm 
  -1 \ge n \ge -l,
\nonumber\\
a_{-l-1}^{\n} &\sim& O(\e^{l}),
\nonumber\\
a_n^{\n} &\sim& O(\e^{\mid n\mid -1}), \hskip 5mm {\rm for } 
\hskip 5mm -l-2 \ge n \ge  -2l,\nonumber\\
a_{-2l-1}^{\n} &\sim& O(\e^{2l-2}),\nonumber\\
a_n^{\n} &\sim& O(\e^{\mid n\mid -3}), \hskip 5mm {\rm for } 
\hskip 5mm    -2l-2 \ge n,
\ena
and for a half-integer $l$,
\bea
a_n^{\n} &\sim& O(\e^{\mid n\mid}), \hskip 5mm {\rm for } 
\hskip 5mm n \ge -(l+\frac12),
\nonumber\\
a_{-(l+\frac12)-1}^{\n} &\sim& O(\e^{(l+\frac12)-1})
\nonumber\\
a_n^{\n} &\sim& O(\e^{\mid n\mid -2}),\hskip 5mm {\rm for } 
\hskip 5mm     -(l+\frac 12)-2 \ge n \ge -2(l+\frac12)-1\nonumber\\
a_{-2(l+\frac12)}^{\n} &\sim& O(\e^{2(l+\frac12)-4}),\nonumber\\
a_n^{\n} &\sim& O(\e^{\mid n\mid -4}), \hskip 5mm {\rm for } 
\hskip 5mm    -2(l+\frac12)-1 \ge n.
\ena
With the above order estimates, we see that how many terms should be 
needed to calculate the coeficients with   the specified accuracy   
of $\e$. 
 
Comming back to $\n$, we assumed that $L_{-2}(\n) \sim O(\e)$ whcih is 
valid if we consider $l \ge \frac 32$. However, this speciality is due to 
the fact that we solved the constraint equation for $n=1$ in Eq.(2.13). Since 
$\n$ is independent of what $n$ we used for solving the constraint equation, 
the result in Eq.(5.2) should 
be valid for all angular momentum case. In fact the result is nonsingular 
for all integer and half-integer values of $l$.

The coefficients $a_n^{\n}$ and also $b_n^{\n}$ up to $O(\e^2)$ (which are 
valid for $l \ge \frac 32$) are obtained 
explicitly by
\bea
a_1^{\n}&=&i{(l+1-s)^2[(l+1)\kappa +imq]\over{2(l+1)^2(2l+1)}}\e\nonumber\\
&&+{(l+1-s)^2\over{2(l+1)^2(2l+1)}}\left[1-i\frac{(l+1)\kappa +imq}
{l(l+1)^2(l+2)} 
mq s^2\right]\e^2+O(\e^3),
\ena
\bea
a_2^{\n}=-{(l+1-s)^2(l+2-s)^2[(l+1)\kappa +imq][(l+2)\kappa +imq]\over
{4(l+1)^2(l+2)(2l+1)(2l+3)^2}}\e^2+O(\e^3)
\ena
\bea
a_{-1}^{\n}&=&i{(l+s)^2[l\kappa -imq]\over{2l^2(2l+1)}}\e\nonumber\\
&&-{(l+s)^2\over{2l^2(2l+1)}}\left[1+i\frac{l\kappa -imq}{(l-1)l^2(l+1)}
mq s^2\right]\e^2+O(\e^3),
\ena
\bea
a_{-2}^{\n}=-{(l-1+s)^2(l+s)^2[(l-1)\kappa -imq][l\kappa -imq]\over
{4(l-1)l^2(2l-1)^2(2l+1)}}\e^2+O(\e^3).
\ena

The coefficients $b_n^{\n}$ are given from Eq.(3.10) by 

\bea
b_1^{\n} = i \frac{(l+1+s)^2}{(l+1-s)^2}a_1^{\n} +O(\e^3),
\ena
\bea
b_2^{\n} = -\frac{(l+1+s)^2(l+2+s)^2}{(l+1-s)^2(l+2-s)^2}a_2^{\n}+O(\e^3),
\ena
\bea
b_{-1}^{\n} = -i\frac{(l-s)^2}{(l+s)^2}a_{-1}^{\n}+O(\e^3),
\ena
\bea
b_{-2}^{\n} = -\frac{(l-1-s)^2(l-s)^2}{(l-1+s)^2(l+s)^2}a_{-2}^{\n}
+O(\e^3).
\ena
By using these coefficients, we can evaluate the ingoing and the 
outgoing amplitudes in infinity. 
From Eq.(4.2), we find by taking $r=0$ that 
$K_{\n} \sim O(\e^{-l+s})$. On the other hand, the estimate of 
$K_{-\n -1}$ needs some care. By taking into account of the 
singular behaviors of gamma functions and the fact that 
the deviation of $\n$ from $l$ starts from the second order of 
$\e$, we find that $K_{-\n -1} \sim O(\e^{l-1+s} \sin i\pi \t)$. 
Thus we obtain
\bea
\frac {K_{-\n -1}}{K_{\n }} \sim O(\e^{2l -1} \sin i\pi \t),
\ena
where $\t=(\e-mq)/\kappa$. In the approximation up to $O(\e^2)$, 
we can neglect $K_{-\n -1}$ term when we restrict $l \ge 3/2$.
We note that for the Schwarzshild case, 
$\t=\e$ so that the ratio in Eq.(5.15) is of order $\e^{2l}$. 

Thus for $l \ge 3/2$, we get the simple expressions for 
the outgoing and the incomming amplitudes as follows;
\bea
A_{out}^{s\,\n}=-ie^{-{i\over{2}}\p(\n+s-i\e)}2^{-1-s+i\e}K_{\n}\sum_{n=-2}
^{2}b_n^{\n}(-i)^n,
\ena
and
\bea
A_{in}^{s\,\n}=ie^{-{i\over{2}}\p(-\n+s-i\e)}2^{-1+s-i\e}K_{\n}
\sum_{n=-2}^{2}b_n^{\n}i^n{\G(n+\n+1-s+i\e)\over{\G(n+\n+1+s-i\e)}}.
\ena
By substituting the coeficients, we can easily calculate the 
amplitudes up to the order $\e^2$. Since the explicit expressions 
are complicated, we present the amplitudes up to $O(\e)$ 
explicitly. We find 
\bea
A_{out}^{s\,\n}&=&-ie^{-{i\over{2}}\p(\n+s-i\e)}2^{-1-s+i\e}K_{\n}
 e^{i[\frac {\kappa}2 (1+\frac {s^2}{l(l+1)})\e +\phi_2 \e^2]
+s[-\frac{mq}{l(l+1)}
\e +\psi_2 \e^2]} \nonumber \\
&&\hskip 2mm \times \left [1+\frac{mqs^2}{2l^2(l+1)^2}\e +d_2 \e^2 \right ],
\ena
and
\bea
A_{in}^{s\,\n}&=&ie^{-{i\over{2}}\p(-\n+s-i\e)}2^{-1+s-i\e}K_{\n}
 e^{i[-\frac {\kappa}2 (1-\frac {s^2}{l(l+1)})\e +\phi_2'\e^2]} 
\nonumber\\
&& \times \left [1+\frac{mqs^2}{2l^2(l+1)^2}\e + d_2 \e^2 \right ],
\ena
where 
\bea
\phi_2&=&\frac{\kappa mq}{2(2l+1)}\left [ 
(l-s)^2 \left \{ \frac {(l-1-s)^2}{2(l-1)l^2(2l-1)} 
-\frac{s^2}{(l-1)l^3(l+1)} \right \} \right.\nonumber\\
 &&\hskip 2mm \left.+\frac 1{4(2l+1)}\left \{ \frac{(l-s)^2}{l}+
\frac{(l+1+s)^2}{l+1} \right \}
\left\{ \frac{(l-s)^2}{l^2}-\frac{(l+1+s)^2}{(l+1)^2} \right \}
\right ]\nonumber\\
&&\hskip 2mm +(l \to -l-1),\\
\psi_2&=&\frac 1{2l+1} \left [ \frac 1 l +(mqs)^2\left \{
\frac {1}{(l-1)l^3(l+1)}+\frac{2l+1}{4l^3(l+1)^3} \right \} \right. 
\nonumber\\
&&\hskip 2mm \left. +\frac{[\kappa(l-1)l-(mq)^2][(l-1)l+s^2]}
{2(l-1)l^2(2l-1)} \right ]
+(l \to -l-1),\\
\phi_2'&=&\frac{mq}{2l+1}\left [ \frac 1 l+ \frac{\kappa s^2(l^2-s^2)}
{2(l-1)l^3(l+1)}+ \frac{\kappa[(l-1)^2-s^2][l^2-s^2]}
{4(l-1)l^2(2l-1)}   \right.\nonumber\\
&&\hskip 2mm  \left. +\frac {\kappa s^2}{8l^2(l+1)^2}
\left \{1-\frac{s^2}{l(l+1)}
   \right \}        \right ] 
+(l \to -l-1),\\
d_2&=& \left[ \frac{(mqs)^2}{4l^2(l+1)^2}-\frac{l^2+s^2}{2l^2(2l+1)}
\left\{1+\frac{(mqs)^2}{(l-1)l^2(l+1)}    \right\} \right. \nonumber\\
&&\hskip 2mm -\frac{[\kappa^2 (l-1)l-(mq)^2][((l-1)^2+s^2)
(l^2+s^2)+4(l-1)ls^2]}{4(l-1)l^2(2l-1)^2(2l+1)}\nonumber\\
&&\hskip 2mm \left. +\frac{\kappa^2}{16}\left\{ 1+\frac{s^2}{l(l+1)}\right 
\}^2\right ]
+(l \to -l-1).
\ena
The above result shows that the absorption coeficient $\Gamma$ in 
Eq.(4.7) is zero up to the order of $\e^2$ for  Kerr black hole.

\sect{Summary and Remarks}
\indent

Analytic solutions of Teukolsky equation are obtained in the form of 
series of hypergeometric functions and Coulomb wave functions. 
The convergence of these solutions are examined. The series solution 
of hypergeometric type is convergent in the region except infinity, 
while the one of Coulomb type is convergent whem $\mid x \mid > 1$. 
The renormalized angular momentua $\n$ turns out to be identical  for 
these two solutions. This fact  enabled us to relate these two 
solutions analytically. 

We examined the $\e$ dependence 
of $a_n^{\n}$ and found that the series corresponds    
essentially to $\e$ except  for some negative integer $n$ 
where  some anomalous behaviors occured for  which we 
need to pay a care to evaluate the coefficients. 
We explicitly calculated $\n$,  the coefficients, 
$A_{out}^{s\n}$ and $A_{in}^{s\n}$ up to the 
order $\e^2$. 

The solutions are not only useful to discuss the low fequency behaviors 
of various physical quantities by applying our solutions, but also 
useful to know the general properties of solutions. For example, we 
consider the $\e$ dependence of the renormalized angular momentum $\n$ 
in the Schwarzschild metrics. $\n$ is determined by solving the 
transcendental equation (2.13) which is composed of $\b_k^{\n}$ and 
$\a_k^{\n}\gamma_{k+1}^{\n}$. These quantities are even functions of 
$\e$ in the Schwarzschild geometry because $\tau=\e$. Therefore, we 
conclude that $\n$ is an even function of $\e$, i.e., $\n(-\e)=\n(\e)$. 
This property is the special one and not valid for the Kerr geometry. 
The fact that the solutions are given by the $\e$ expansion is important 
because the $\e$ expansion corresponds to the Post-Minkowskian $G$ expansion 
and also to the post-Newtonian expansion when they are applied to the 
gravitational radiation from a particle in circular orbit around a 
black hole. The solutions can be used for the analysis of the gravitational 
radiation from coalescing compact binary systems. Since the analytical 
properties and the convergences are known, the solutions will give a 
powerful method for numerical computation and will contribute 
to construct the theoretical template towards the gravitational observation 
by LIGO and VIRGO projects.

\vskip 4cm
{\Huge Acknowledgment}

We would like to thank to M. Sasaki for comments and 
encouragements. This work is supported in part by 
the Japanese Grant-in-Aid for Science Reserch of
Ministery of Education, Science, Sports and Culture, 
No. 06640396.

\newpage

\setcounter{section}{0}
\renewcommand{\thesection}{\Alph{section}}
\newcommand{\apsc}[1]{\stepcounter{section}\noindent
\setcounter{equation}{0}{\Large{\bf{Appendix\,\thesection:\,{#1}}}}}

\apsc{Spheroidal Teukolsky equation}

\indent

 To expand radial Teukolsky equation, we have to derive the eigenvalue of 
the spheroidal Teukolsky equation. Fortunately, our method is also 
available in expansion of spheroidal Teukolskuy equation. Fackerell 
expand spheroidal Teukolsky equation in terms of Jacobi 
functions\cite{Fackerell}. In our expansion method, 
we can derive the eigenvalue of the spheroidal field equation 
which appear in the radial Teukolsky equation.\\
The separated spheroidal Teukolsky equation is
\bea
\left[(1-x^{2}){d^2\over{dx^{2}}}-2x{d\over{dx}}+{\xi}^{2} x^{2}
-{m^{2}+s^{2}-2msx\over{1-x^{2}}}-2s\xi x+E \right] S(x)=0,
\ena
where $\xi = a\,\z, x = cos\theta$.\\
We make transformation as
\bea
S(x)=e^{\xi x}\left({1-x\over{2}}\right)^{\a}
\left({1-x\over{2}}\right)^{\b}u(x),
\ena
where $\a = |m+s|, \b = |m-s|$,\\
then we recast the equation:
\bea
&&(1-x^{2})u''+\left[\b-\a-(2+\a+b)x\right]u'+
\left[E-{\a+\b\over{2}}\left({\a+\b\over{2}}+1\right)\right]u \nonumber\\
&=&\xi\left[-2(1-x^{2})u'+(\a+\b+2s+2)xu-(\xi+\b-\a)u\right].
\ena
As a solution of first order equation, we use Jacobi functions which is 
the solution of the equation
\bea
(1-x^{2})U_{n}^{(\a,\b)}{}''+\left[\b-\a-(\a+\b+2)x\right]U_{n}^{(\a,\b)}{}'+
n(n+\a+\b+1)U_{n}^{(\a,\b)}=0.
\ena
By using recursion relations of Jacobi functions, we can analytically expand 
$u$ in terms of $U_{n}^{(\a,\b)}$\cite{Fackerell},
\bea
u_{n}=\sum_{j=-\infty}^{\infty}c_{j}\,U_{n+j}^{(\a,\b)},
\ena
where $l=n+(\a+\b)/2$.\\
We can expand $c_{j}, E$ in $\xi$:
\bea
c_{j}=\sum_{k=0}^{\infty}c_{j}^{(k)}\xi^{k}, \qquad
E=\sum_{k=0}^{\infty}E^{(k)}\xi^{k},
\ena
here we set $c_{0}=1,c_{n\neq0}^{(0)}=0$ and $E^{(0)}=l(l+1)$ .\\
\bea
c_{1}&=&{(2l+2)^2-(\a+\b)^2\over{(2l+1)(2l+2)^2}}(l+s+1)\xi,
\ena
\bea
c_{-1}&=&{(2l)^2-(\a-\b)^2\over{(2l)^2(2l+1)}}(l-s)\xi,
\ena
\bea
c_{2}&=&{[(2l+4)^2-(\a+\b)^2][(2l+2)^2-(\a+\b)^2]
\over{4(2l+1)(2l+2)^2(2l+3)^2(2l+4)}}(l+s+1)(l+s+2)\xi^2,
\ena
\bea
c_{-2}&=&{[(2l-2)^2-(\a-\b)^2][(2l)^2-(\a-\b)^2]
\over{4(2l-2)^2(2l-1)(2l)(2l+1)^2}}(l-s)(l-s-1)\xi^2
\ena
and $E$ in Eq.(1.3) which  is identical to that of Fackerell\cite{Fackerell} 
who also has shown that the convergency of the expansion in terms of 
Jacobi functions.\\

\apsc{Derivations of equations (2.2) and (3.2)}

\indent

The radial Teukolsky equation is written by using the variable 
$y=\z r$ with $y_+=\z r_+$ and  $y_-=\z r_-$ as
\bea
&&\frac {d^2 R}{dy^2} + (s+1)(\frac 1{y-y_+}+\frac 1{y-y_-})
\frac{dR}{dy}\nonumber\\
&&\hskip 2mm +[1+\frac1{y-y_+}(\e+is+\frac{\e+2is}{\kappa }) +
   \frac1{y-y_-}(\e+is - \frac{\e+2is}{\kappa })\nonumber\\
&&\hskip 5mm +\frac 1{(y-y_+)^2}\frac{(\e-is+\t)^2+s^2}{4} +
   \frac1{(y-y_-)^2}\frac{(\e-is-\t)^2+s^2}{4} \nonumber\\
&&\hskip 5mm +\frac 1{(y-y_+)(y-y_-)}(-\lambda +\frac {\e^2}2 -
    \frac{\t^2}2 -i\e s -\e mq)]R = 0.
\ena

\noindent
(a) The expansion in terms of hypergeometric functions

We define a new variable $x$ by 
\bea
y-y_+=\e \kappa (-x),\qquad y-y_-=\e \kappa (1-x).
\ena
We rewrite $R$ as
\bea
R=(-x)^{\a}(1-x)^{\b}\tilde R.
\ena
in order to eliminate the terms proportional to $1/x^2$ and $1/(1-x)^2$. 
Then, $\a$ and $\b$ are determined to be one of following values, 
\bea
\a_{\pm}=\frac 12 (-s \pm i(\e-is+\t)),\qquad 
\b_{\pm}=\frac 12 (-s \pm i(\e-is-\t)),
\ena
respectively. 
With these choices of $\a$ and $\b$ and  the change of $\tilde R$ in the 
following form
\bea
\tilde R = e^{i\e \kappa  x}p,
\ena
the equation for $p$ is expressed by 
\bea
 &&x(1-x)p''+[(2\a+s+1)-2(\a+\b+s+1)x]p'-abp\nonumber\\
&&\hskip 2mm =-2i\e \kappa  x(1-x)p' +2i\e \kappa (\a+\b+1+i\e)xp
+[-\lambda+2\a\b+(s+1)(\a+\b) 
   \nonumber\\
&& \hskip 2mm -i\e \kappa (2\a+s+1)+\e \kappa (\e+is)+\frac32 \e^2-
\frac12 \t^2+i\e s -\e mq-ab]p,
\ena
where $a$ and $b$ are chosen such that the equality 
\bea
a+b=2(\a+\b)+2s+1
\ena
is satisfied and also they take some  simple forms. 

If we take one of choices $(\a_-,\b_+)$ and $(\a_-,\b_-)$, 
the form of $R$ defined in 
Eq.(B.3) becomes a suitable form for the solution which satisfies the incoming
boundary condition on the outer horizon. On the other hand, the choice of 
$(\a_+,\b_-)$ or $(\a_+,\b_+)$ is suitable to obtain the solution which 
satisfies the outgoing boundary condition. In the text, we took  
$(\a_-,\b_+)$ for the solution satisfying the incoming boundary 
condition in which case the 
above equation (B.6) reduces to the one in Eq.(2.2), by taking $a=\n+1-i\t, 
b=-\n-i\t$. 
For the solution satisfying the outgoing boundary condition, we took  
 $(\a_+,\b_-)$ in which case the equation becomes a similar form.

\noindent
(b) The expansion in terms of Coulomb wave functions

We  take the parameterization 
\bea
R=[(y-y_+)(y-y_-)]^{-(1+s)/2}\left ( \frac{y-y_+}{y-y_-} \right )^\gamma f
\ena
and determine $\gamma$ to eliminate the singularity proportional to 
$1/(y-y_-)^2$. Then we find $\gamma$ should take one of the following values
\bea
\gamma_\pm=\frac 12 [-1 \pm i(\e -is-\t)].
\ena
If we take one of these values of $\gamma$, the equation for $f$ 
becomes with $z=y-y_+=-\e \kappa  x$
\bea
z^2f''+[z^2+2(\e+is)z]f
&=&-\e \kappa z(f''+f)-2\a \e \kappa  f'-{\e \kappa [2\gamma -\t(\e -is)]
\over{z}}f\nonumber\\
&&+[\lambda+s(s+1)-2\e^2 +\e mq -\e \kappa (\e+is)]f.
\ena
The equation (3.2) in the text is obtained by taking $\gamma=\gamma_-$. 
The choice $\gamma=\gamma_-$ gives 
a Coulomb type 
solution which matches with the hypergeometric type solution 
with $(\a_-,\b_+)$ as we 
saw in Eq.(4.1). We can also obtain the solution by choosing $\gamma_+$ 
which matches with the hypergeometric one with $(\a_-,\b_-)$.\\

\end{document}